\documentclass[sigconf]{acmart}

\usepackage{booktabs} 
\usepackage{siunitx}

\AtBeginDocument{%
  }

\copyrightyear{2025}
\acmYear{2025}
\setcopyright{acmlicensed}\acmConference[SIGSPATIAL '25]{The 33rd ACM International Conference on Advances in Geographic Information Systems}{November 3--6, 2025}{Minneapolis, MN, USA}
\acmBooktitle{The 33rd ACM International Conference on Advances in Geographic Information Systems (SIGSPATIAL '25), November 3--6, 2025, Minneapolis, MN, USA}
\acmDOI{10.1145/3748636.3762720}
\acmISBN{979-8-4007-2086-4/2025/11}

\begin{document}

\title{Mapping the Storm: Geospatial Impacts of Severe Weather on LEO Network Performance}

\author{Sina Ehsani}
\email{se@armada.ai}
\orcid{0000-0002-6009-7612}
\affiliation{%
  \institution{Armada}
  \city{San Francisco}
  \state{California}
  \country{USA}
}

\author{Bhanu Pallakonda}
\email{bp@armada.ai}
\affiliation{%
  \institution{Armada}
  \city{San Francisco}
  \state{California}
  \country{USA}
}


\author{Pragyana K. Mishra}
\email{pm@armada.ai}
\affiliation{%
  \institution{Armada}
  \city{San Francisco}
  \state{California}
  \country{USA}
}


\begin{abstract}
   Low Earth Orbit (LEO) satellite constellations, led by deployments such as Starlink, are playing an increasingly pivotal role in enabling global broadband connectivity. However, the reliability and performance of these space-based networks are highly sensitive to environmental dynamics, particularly localized weather phenomena that exhibit strong spatio-temporal variability. In this study, we present a continental-scale geospatial analysis of weather-induced performance degradation in the Starlink LEO network, with a focus on the contiguous United States.
   
   Leveraging a unique dataset comprising more than 870,000 terminal hours of minute-level telemetry from 1,292 Starlink terminals collected between February and April 2025, we integrate high-resolution localized weather observations to quantify the impact of various meteorological conditions. We evaluated key performance indicators (KPIs)—including ping latency, ping drop rate, and signal quality—using spatial join techniques and time-aligned correlation with classified weather events.
   
   Our analysis reveals that severe weather events, such as thunderstorms with heavy rain or snow, have a pronounced effect on network performance. In particular, more than 55\% affected terminals experienced substantial degradation (signal quality $< 0.5$ or ping drop rate $> 0.5$), with some conditions such as 'thunderstorm with heavy rain' exceeding a 70\% disruption rate. Temporal continuity analysis at the minute level shows that such degradation can lead to sustained impairments or full service outages lasting from several minutes to multiple hours.
   
   This work contributes to the first large-scale empirical study linking LEO satellite Internet performance with fine-grained weather data in both space and time. Our findings offer actionable insights for geospatial predictive modeling, weather-aware network provisioning, and resilient satellite communication system design. We also propose a framework for incorporating weather-inferred performance variability into future geospatial planning and service-level forecasting tools for LEO-based Internet systems.
\end{abstract}



\begin{teaserfigure}
  \includegraphics[width=\textwidth]{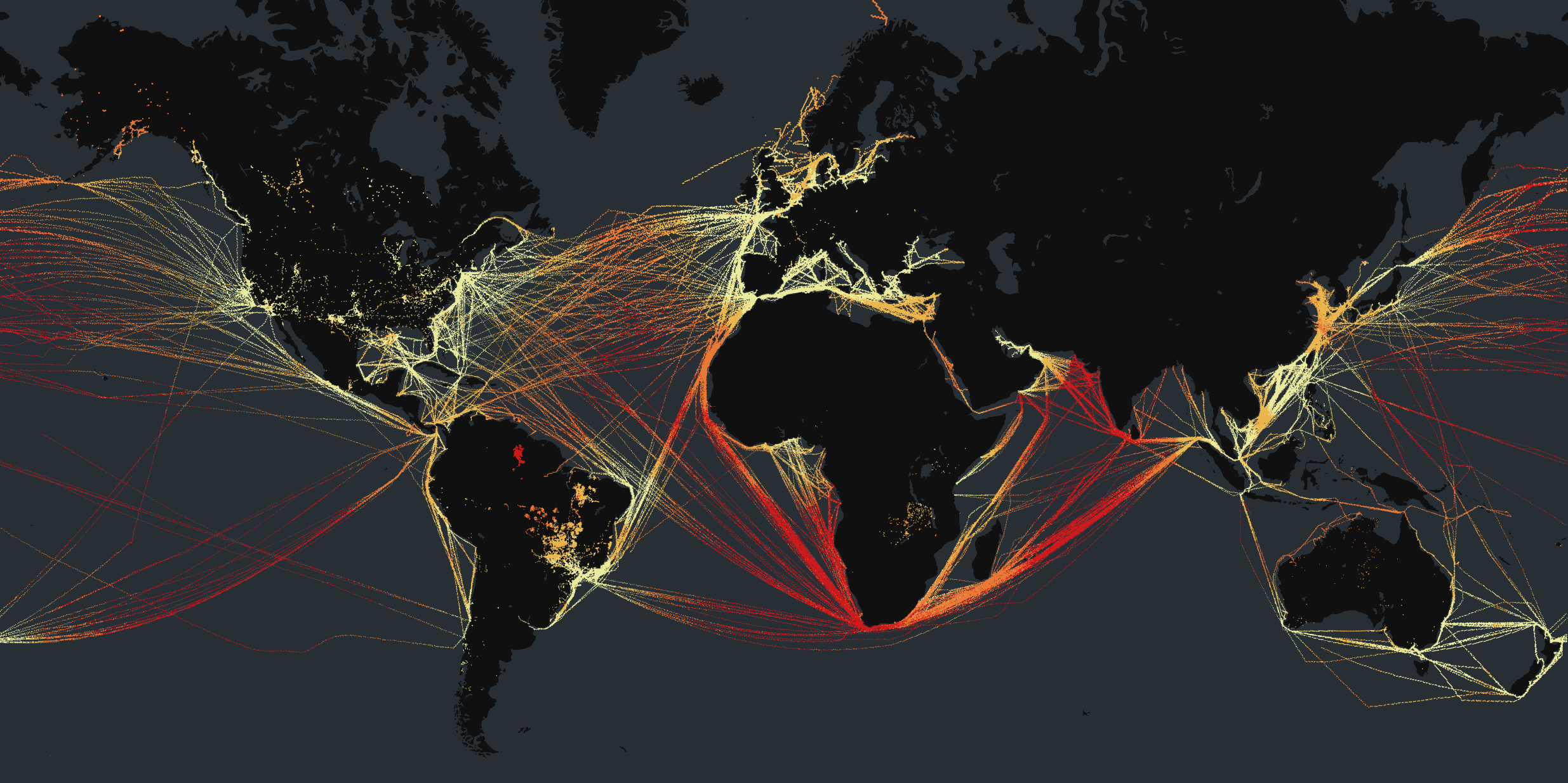}
  \caption{Global distribution of a subset of Starlink terminals managed by Armada Edge Platform (AEP): average latency visualized per H3 cell (higher latency indicated in red)}
  \Description{A fleet map of all the Starlink terminals managed by Armada, color coded by their latency.}
  \label{fig:teaser}
\end{teaserfigure}


\maketitle
\section{Introduction}
The deployment of large-scale Low Earth Orbit (LEO) satellite constellations, led by systems such as SpaceX Starlink, OneWeb, Amazon Kuiper, and Telesat, is transforming global broadband connectivity. Unlike traditional geostationary systems, LEO networks leverage thousands of rapidly orbiting satellites at altitudes between 500–1,200 km to deliver low-latency, high-throughput connectivity, particularly targeting remote and underserved regions. With Ku and Ka-band radio frequencies enabling gigabit-speed user links, and inter-satellite laser links (ISLLs) minimizing backhaul dependency \citep{hasan2016ka, del2019technical}, As of February 2025 Starlink supports over 5 million subscribers globally and serves as a testbed for the broader viability of orbital mesh networking.

Despite these technological advancements, the physical-layer reliability of LEO internet remains vulnerable to the spatio-temporal dynamics of the Earth’s atmosphere. Atmospheric phenomena--including rain, snow, cloud water, ice crystals, and ionospheric scintillation--introduce non-stationary, localized impairments to radio frequency (RF) propagation \citep{sharma2024doppler, sterenborg2005scintillation, sonth2025amplitude}. Of particular concern is rain fade \citep{omotosho2010study}, which causes significant attenuation in the Ku-band (12–18 GHz), a frequency regime where signal loss correlates nonlinearly with rainfall intensity, rate, and droplet size distribution \citep{ullah2025impact, cakaj2009rain}. Additional disruptions stem from snow accumulation on user terminals \citep{freyer2014combating}, cloud cover \citep{ullah2025impact}, gaseous absorption \citep{zubair2011atmospheric}, scintillation effects in the ionosphere and troposphere \citep{bonora2025performance, ma2024integrated}, tropospheric multi-path interference, and antenna beam misalignment during storms. 
These effects are highly spatially heterogeneous and temporally transient, posing significant challenges to both performance prediction and quality-of-service assurance in live deployments.

In response to these challenges, the scientific community has progressively worked to quantify their impacts on LEO systems like Starlink. Studies have documented performance degradation due to rain, cloud cover, and thunderstorms, primarily in localized studies \citep{ullah2025impact, ahmad2025geomagnetic}. This research, complemented by broader work in spatio-temporal network and geospatial data analysis \citep{goodge2025spatio}, offers valuable insights. While the RF and networking research communities have made progress in characterizing atmospheric effects on satellite communication \citep{mertens2023nairas, kalaivaanan2020evaluation, parker2025greenhouse}, existing studies have several critical limitations. First, most prior analyses are either simulation-based or limited to single-point measurements, lacking both spatial generalizability and empirical scale \citep{ullah2025impact, ahmad2025geomagnetic}. Second, they tend to focus on aggregate throughput or latency, without considering fine-grained quality metrics such as packet loss rate or terminal-level signal integrity. Third, they often omit minute-level temporal resolution, which is essential for detecting bursty, short-lived yet impactful disruptions common in real-world weather systems.

To address these gaps, this paper presents the first continental-scale spatio-temporal study quantifying the impact of severe weather on LEO broadband performance, using Starlink as the operational platform. We analyze a dataset of 870,000 terminal-hours from 1,292 geographically distributed Starlink terminals, recorded at minute-level granularity between February and April 2025. This performance data, primarily aggregated to hourly intervals, is joined with high-resolution, localized weather observations, enabling a robust analysis of meteorological impacts. Our methodology also incorporates deep-dive case studies using minute-level data to capture the fine-grained dynamics of service degradation.

Our research provides three key contributions:
\begin{enumerate}
\item  We provide statistically robust evidence quantifying the degradation in Starlink KPIs—including ping latency, ping drop rate, and signal quality—under varying weather conditions. Our findings show that over 55\% of terminals affected by severe weather exhibit significant performance issues, with "Thunderstorm with heavy rain" associated with such issues in over 70\% of cases.

\item Through minute-resolution case studies, we characterize the severity and duration of service impairments. We show that in notable severe weather episodes, disruptions can persist for multiple minutes to over an hour, with clear patterns of degradation and recovery correlating with storm dynamics.

\item We provide a foundation for weather-aware network provisioning and predictive service degradation modeling, offering utility for both LEO operators and geospatial data scientists. Our results highlight the need for spatio-temporal forecasting models that incorporate atmospheric uncertainty into network resilience planning.
\end{enumerate}

A central finding of this study is that the Starlink network exhibits high baseline stability. Across our dataset, over 90\% of terminal-hours showed no significant disturbances in either ping drop rate or signal quality. This general reliability, however, is markedly challenged by specific, severe weather events. While most moderate weather has a limited impact, this paper will demonstrate that severe conditions can dramatically increase the likelihood of performance degradation. This research sits at the intersection of wireless networking, geospatial analytics, and environmental informatics, and extends existing work on spatially grounded QoS modeling \citep{huang2022fine, zou2022deeptsqp} by contributing a large-scale empirical analysis of network resilience under environmental uncertainty.

The remainder of this paper is organized as follows: Section \ref{sec:methodology} details our data sources, preprocessing steps, and analytical methodology. Section \ref{sec:results} presents the results of our spatio-temporal analysis. Section \ref{sec:discussion_future} discusses the implications of our findings, limitations of our approach, and future research directions in predictive modeling and geospatial network resilience.

\section{Data and Methodology}
\label{sec:methodology}
This study adopts a large-scale, data-driven methodology to investigate the spatio-temporal relationship between localized weather phenomena and Starlink LEO network performance. Our approach involves the systematic collection, spatial indexing, temporal aggregation, and statistical analysis and correlation of satellite terminal telemetry with fine-grained environmental data.

\subsection{Starlink Performance Data Acquisition and Processing}
\label{ssec:starlink_data}
Performance telemetry from Starlink user terminals are continuously collected via Armada’s Atlas platform, which manages a distributed fleet of LEO-connected assets worldwide. Atlas interfaces directly with Starlink terminal APIs, continuously ingesting fine-grained operational metrics at sub-minute resolution. These raw streams include device-level observations such as ping latency, ping drop rate, and terminal-reported signal quality, all of which serve as core Key Performance Indicators (KPIs) in our analysis. It is important to note that this terminal-side telemetry provides user-experienced performance indicators but does not include internal network state information from the provider, such as satellite handover commands or real-time satellite position data, which are outside the scope of this observational study.

Given the high frequency and scale of data across thousands of distributed terminals, we implemented a daily ETL pipeline that performs temporal aggregation into hourly intervals. For each terminal-hour, the system computes the mean and standard deviation of each KPI. The resulting aggregates are persisted in Delta Lake tables, enabling scalable analytical queries and efficient I/O performance on distributed compute engines.

 
A crucial aspect of our data processing involves tracking the geographic location of each terminal. Since a subset of the managed terminals are mobile (e.g., installed on vehicles or marine assets), the last known geographic coordinates (latitude and longitude) of each terminal are updated and recorded with each hourly aggregation. For mobile terminals, the location at the start of the hourly aggregation period was used for querying weather conditions, under the assumption that weather conditions influencing satellite link performance are relatively consistent within that hour for the terminal's immediate vicinity. The H3 hexagonal hierarchical spatial index \citep{uber2018h3} is utilized to associate terminal locations with specific geographic grid cells, which aids in visualizing spatial distributions, such as the terminal density map presented in Figure \ref{fig:teaser}.

\begin{figure}[t]
  \centering
  \includegraphics[width=\linewidth]{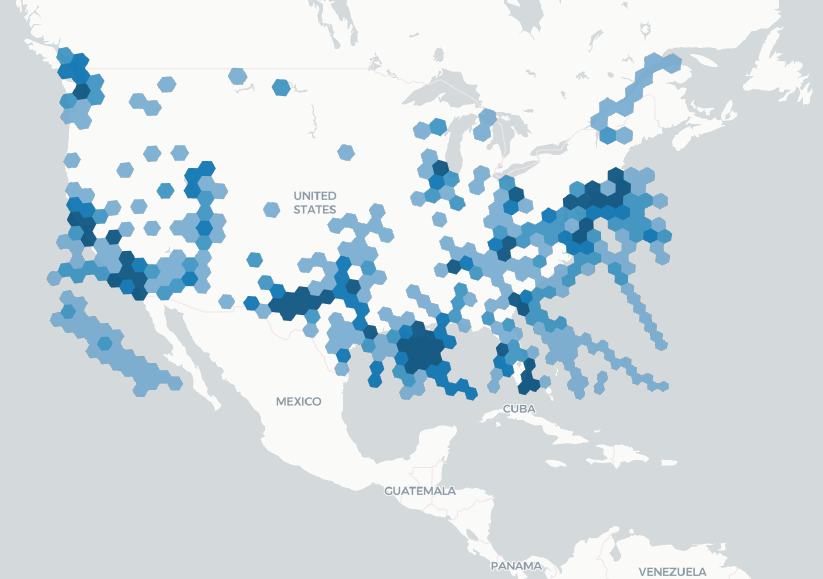} 
  \caption{The geographic distribution of the 1,292 Starlink terminals included in this study spans the continental United States as of April 2025. The map uses H3 level 3 hexagonal cells, color-coded to represent the number of unique terminals per cell, with darker shades indicating higher concentrations—up to a maximum of 32 terminals in a single cell. This visualization highlights both the spatial coverage and density variation of the dataset across the study region.}
  \label{fig:terminal_distribution_map}
\end{figure}

For this study, we focus on a cohort of 1,292 unique Starlink terminals. The scope of this study is limited to terminals operating within the continental United States, thereby excluding terminals in Alaska, Hawaii, and other US territories for this specific analysis. The geographic distribution of these terminals is illustrated in Figure \ref{fig:terminal_distribution_map}, highlighting the spatial coverage of our dataset across the continental U.S. with varying densities. The analysis period spans from February 24, 2025, to April 2, 2025. After data cleaning (detailed in Section \ref{ssec:data_merging_preprocessing}), this yielded approximately 870,000 unique terminal-hour data records, where both the performance and corresponding weather data were successfully obtained and completed. The primary Starlink KPIs considered in this study are detailed in Table \ref{tab:key_variables}.

\subsection{Weather Data Acquisition}
\label{ssec:weather_data}
Localized weather data for each Starlink terminal is retrieved using the Weatherbit API \citep{weatherbit2025}, a commercial service that offers high-resolution historical and forecast meteorological data. For every hourly performance record of a Starlink terminal, a corresponding historical weather snapshot is obtained by querying Weatherbit based on the terminal's last known geographic coordinates (latitude and longitude) for that specific hour. The API returns a rich set of atmospheric variables at hourly granularity, including but not limited to temperature, precipitation intensity, cloud cover percentage, and solar angles. These parameters are essential for evaluating environmental conditions that influence signal propagation and terminal performance. The primary Starlink Key Performance Indicators (KPIs), along with the selected weather variables relevant to our spatio-temporal analysis, are enumerated in detail in Table \ref{tab:key_variables}. This study uses these ground-level observations as a proxy for the atmospheric conditions along the full slant path from the terminal to the satellite. While factors like satellite elevation angle affect this path's length through the atmosphere, this approach is a common and necessary simplification for large-scale geospatial analyses where precise, real-time 3D atmospheric modeling is not feasible.

\begin{table*}[htpb!]
\centering
\caption{Primary Starlink Key Performance Indicators (KPIs) and Weather Variables Used in Analysis.}
\label{tab:key_variables}
\begin{tabular}{@{}llll@{}}
\toprule
\textbf{Variable Group} & \textbf{Parameter Name} & \textbf{Description} & \textbf{Unit} \\
\midrule
\multicolumn{4}{@{}l}{\textit{Starlink Performance KPIs (Hourly Aggregates)}} \\
 & Ping Latency & Average round-trip time (RTT) to Starlink Point of Presence (PoP) & ms \\
 & Ping Drop Rate & Average percentage of pings to PoP not receiving a response & \% \\
 & Signal Quality & Terminal-reported Signal-to-Noise Ratio (SNR), normalized & 0--1 \\
 & Downlink Throughput & Average data rate from satellite to terminal & Mbps \\
 & Uplink Throughput & Average data rate from terminal to satellite & Mbps \\
 & Obstruction Percent Time & Percentage of time sky view was reported as obstructed & \% \\
\addlinespace
\multicolumn{4}{@{}l}{\textit{Weather Variables (from Weatherbit API, Hourly)}} \\
 & Temperature & Ambient air temperature & $^{\circ}$C \\
 & Dew Point & Dew point temperature & $^{\circ}$C \\
 & Relative Humidity & Relative humidity & \% \\
 & Cloud Coverage & Percentage of sky covered by clouds & \% \\
 & Precipitation Rate & Liquid equivalent precipitation rate & mm/hr \\
 & Snowfall Rate & Snowfall accumulation rate & mm/hr \\
 & Wind Speed & Average wind speed & m/s \\
 & Weather Description & Textual summary of weather conditions & Text \\
 & Solar Azimuth Angle & Horizontal angle of the sun relative to true north & degrees \\
 & Solar Elevation Angle & Vertical angle of the sun above the horizon & degrees \\
\bottomrule
\end{tabular}
\end{table*}

\subsection{Data Merging and Preprocessing for Analysis}
\label{ssec:data_merging_preprocessing}
To construct a unified analytical dataset, the hourly aggregated Starlink performance records - comprising the mean and standard deviation of the KPIs along with geolocation metadata - were joined with hourly localized weather observations based on terminal ID, timestamp, and spatial location. This join yields a comprehensive spatio-temporal dataset, where each row represents one hour of operation for a specific Starlink terminal, enriched with prevailing environmental conditions. The variables used in this analysis, including both performance metrics and atmospheric attributes, are detailed in Table \ref{tab:key_variables}.

To ensure the reliability and interpretability of the results, a multi-stage data cleaning and filtering process was applied:
\begin{enumerate}
\item Geographic Scope Filtering: Terminal records outside the continental United States (CONUS) were excluded to maintain consistency in weather data availability and satellite visibility zones. This excluded installations in Alaska, Hawaii, and U.S. territories.

\item Missing Data Elimination: Records with null or incomplete values in any critical field—such as KPI metrics (e.g., signal quality or ping drop rate), terminal geolocation, or primary weather attributes—were discarded. This step ensures data integrity for correlation and statistical inference.

\item Obstruction Filtering: To isolate performance degradation primarily attributable to transient weather phenomena rather than persistent local obstructions, we identified terminals with high sustained obstruction rates. Specifically, terminals with an average ObstructionPercentTime > 30\% over the entire study period were removed. This metric reflects the fraction of time the terminal’s field of view was occluded, typically by static physical impediments like trees or buildings. Only two terminals met this exclusion threshold, suggesting that persistent obstruction was rare but important to control for in the broader analysis.
\end{enumerate}

Following these steps, the final analytical dataset consisted of approximately 870,000 terminal-hour records derived from 1,292 unique terminals. This dataset is characterized by spatial diversity, temporal coverage across multiple weather systems, and metric completeness, making it well-suited for robust statistical and geospatial analysis.

Importantly, no additional filtering or outlier removal based on the magnitude of performance degradation was performed. This preserves genuine variability in network behavior, particularly under severe weather conditions, ensuring that the dataset captures real-world edge cases, including full service outages and periods of extreme degradation.

\subsection{Definition of Severe Weather Events and Performance Issues}
\label{ssec:definitions}
To specifically investigate the impact of more extreme conditions, we categorized certain weather descriptions from the Weatherbit API as "Severe Weather Events." For this study, these include: "Thunderstorm with rain", "Thunderstorm with heavy rain", "Thunderstorm with light rain", "Heavy rain", "Heavy snow", and "Heavy sleet" as these categories represent conditions known to cause significant Ku-band signal attenuation or combined physical effects (e.g., antenna wetting and signal fade).

A "Performance Issue" during an hourly interval was defined if, for that hour, either the average Signal Quality dropped below $0.5$ (on its $0-1$ normalized scale, where 1 is optimal) or the average Ping Drop Rate exceeded $0.5$ (i.e., $50\%$). These thresholds were selected to identify periods of substantial service degradation. For context:
\begin{itemize}
    \item A Ping Drop Rate exceeding 5\% is generally considered problematic for many internet applications, with rates approaching 50\% indicative of a near-unusable connection for most purposes \citep{mansfield2009computer, sommers2005improving}.
    \item While specific Signal-to-Noise Ratio (SNR) thresholds vary by system, a signal quality metric dropping to less than half of its optimal value typically signifies a severely compromised link with high bit error rates and instability \citep{starlink_enterprise_getting_started}. 
    \item Empirically, within our dataset, these chosen thresholds represent significant deviations (typically exceeding three standard deviations from the mean observed during clear weather, non-obstructed conditions) from baseline performance, aligning with periods where service would be critically impacted.
\end{itemize}

While the primary aggregated analysis uses these hourly KPI thresholds, qualitative assessments of 'total disconnect' periods for individual case studies (presented in Section \ref{ssec:results_deep_dives}) were based on observations of minute-level signal quality reporting as zero or sustained periods where no telemetry was received from the terminal. These 'total disconnect' instances are highlighted in case studies to illustrate the most severe impacts but are not systematically quantified across the entire dataset in the current aggregated impact assessment.

\subsection{Analytical Methods}
\label{ssec:analytical_methods}
Our analysis employs several quantitative and qualitative methods to investigate the spatio-temporal relationship between weather and Starlink performance.

\begin{enumerate}
    \item \textbf{Descriptive Spatio-Temporal Statistics:} As a first step, we calculated descriptive statistics (mean, standard deviation) for all primary KPIs and weather variables to establish a baseline understanding of the dataset (Table \ref{tab:descriptive_stats}). Geospatial visualization using an H3 grid map was employed to illustrate the geographic distribution of the terminal cohort (Figure \ref{fig:terminal_distribution_map}).
    \item \textbf{Correlation and Trend Analysis:} Pearson correlation coefficients \citep{pearson1896vii} were calculated between key Starlink KPIs and selected continuous weather variables. To identify statistically significant linear relationships, p-values were computed, with a significance level of $\alpha = 0.05$. For visualizing trends in scatter plots (e.g., Figure \ref{fig:temp_latency_linear}), we utilized Locally Estimated Scatterplot Smoothing (LOESS) \citep{cleveland1979robust} to illustrate local patterns without assuming a global linear model.
    \item \textbf{Comparative Analysis by Weather Category:} To compare Starlink performance across different discrete weather categories, we used an Analysis of Variance (ANOVA) \citep{fisher1970statistical}. The assumption of homogeneity of variances was assessed using Levene's test \citep{levene1960robust}. Following a significant overall ANOVA result, pairwise comparisons between weather categories were conducted using Tukey's Honestly Significant Difference (HSD) post-hoc test \citep{tukey1949comparing} to identify which specific groups differed significantly from one another.
    \item \textbf{Threshold-Based Impact Assessment:} The percentage of terminal-hours experiencing "Performance Issues" (as defined in Section \ref{ssec:definitions}) was calculated for different weather categories. To determine if the likelihood of an issue was statistically dependent on weather severity, we used the Chi-squared ($\chi^2$) test of independence \citep{pearson1900x} on a 2x2 contingency table (Severe vs. Non-Severe weather against Issue vs. No Issue).
    \item \textbf{Temporal Case Study Analysis:} For selected severe weather events, minute-level Starlink performance data from individual terminals was analyzed. This involved plotting time-series data to qualitatively assess the fine-grained temporal dynamics of service degradation, including the duration of outages or periods of severely impaired quality, providing deeper context to the aggregated hourly results.
\end{enumerate}

\section{Results}
\label{sec:results}
This section presents the empirical findings from our spatio-temporal analysis of weather impacts on Starlink performance. We begin with an overview of the dataset characteristics, followed by an examination of general correlations between weather variables and Starlink Key Performance Indicators (KPIs). We then delve into the specific impacts of different weather categories, particularly severe weather events, on service quality and conclude with illustrative case studies based on minute-level analysis.

\subsection{Dataset Overview and Descriptive Statistics}
\label{ssec:results_descriptive}
The final analytical dataset, after preprocessing and filtering as described in Section \ref{ssec:data_merging_preprocessing}, consists of approximately 870,000 terminal-hour records from 1,292 unique Starlink terminals located across the continental United States during the period of February 24, 2025, to April 2, 2025. The geographic distribution of these terminals is shown in Figure \ref{fig:terminal_distribution_map}.

Table \ref{tab:descriptive_stats} provides summary statistics for the primary Starlink KPIs and key environmental variables across all terminal-hour records in the dataset. These statistics offer a baseline understanding of the typical performance metrics and experienced weather conditions during the study period. For example, the overall mean ping latency was \SI{28.23}{\milli\second}. Notably, Signal Quality remained consistently high (mean 0.976, std. dev. 0.059 on a 0--1 scale), indicating a generally robust satellite link. Ping Drop Rate was also very low on average (mean \SI{0.024}{\percent}), although its standard deviation (\SI{0.064}{\percent}) suggests some inherent variability, likely encompassing sporadic instances of packet loss that become more systematically prevalent under adverse weather conditions. This overall baseline of high signal quality and typically low packet loss underscores the significance of deviations from this stability, which are explored in detail in subsequent sections focusing on specific weather impacts.  It is important to note that the reported mean throughput values (Downlink: \SI{2.41}{\mega\bit\per\second}, Uplink: \SI{0.58}{\mega\bit\per\second}) represent averages over all hourly intervals, including periods of low or no user activity, and thus may appear lower than typical peak performance rates often associated with the service.

\begin{table}[t]
\centering
\caption{Overall Descriptive Statistics for Key Starlink KPIs and Weather Variables (N $\approx$ 870,000 terminal-hours).}
\label{tab:descriptive_stats}
\sisetup{round-mode=places,round-precision=3} 
\begin{tabular}{@{}l S[table-format=3.3] S[table-format=3.3]@{}} 
\toprule
\textbf{Parameter} & {\textbf{Mean}} & {\textbf{Std. Dev.}} \\ 
\midrule
\multicolumn{3}{@{}l}{\textit{Starlink Key Performance Indicators (KPIs)}} \\
Ping Latency (\si{\milli\second}) & 28.2264 & 5.3628 \\
Ping Drop Rate (\si{\percent}) & 0.0243 & 0.0643 \\
Signal Quality (0--1) & 0.9765 & 0.0592 \\
Downlink Throughput (\si{\mega\bit\per\second}) & 2.4135 & 7.4061 \\
Uplink Throughput (\si{\mega\bit\per\second}) & 0.5791 & 1.5415 \\
Obstruction Time (\si{\percent}) & 0.3529 & 2.0226 \\
\addlinespace
\multicolumn{3}{@{}l}{\textit{Key Weather Variables}} \\
Temperature (\si{\degreeCelsius}) & 15.4181 & 7.8952 \\
Dew Point (\si{\degreeCelsius}) & 6.2133 & 9.6835 \\
Relative Humidity (\si{\percent}) & 60.8660 & 24.9853 \\
Cloud Coverage (\si{\percent}) & 38.1818 & 38.4044 \\
Precip. Rate (\si{\milli\meter\per\hour}) & 0.0863 & 0.7719 \\
Snowfall Rate (\si{\milli\meter\per\hour}) & 0.0312 & 0.6760 \\
Wind Speed (\si{\meter\per\second}) & 4.1641 & 2.6574 \\
Solar Azimuth (\si{\degree}) & 178.8336 & 95.6067 \\
Solar Elevation (\si{\degree}) & 2.0493 & 37.8099 \\
\bottomrule
\end{tabular}
\end{table}

The dataset captured a diverse range of weather conditions, with "Clear Sky" and various cloud cover types being the most prevalent, collectively accounting for approximately 94\% of the observed terminal-hours. Severe weather events, as defined in Section \ref{ssec:definitions}, constituted a smaller but significant portion of the dataset, allowing for focused analysis of their impacts.

\subsection{General Correlations Between Weather Variables and Starlink KPIs}
\label{ssec:results_correlations}
\begin{figure}[t]  
 \centering
 \includegraphics[width=\linewidth]{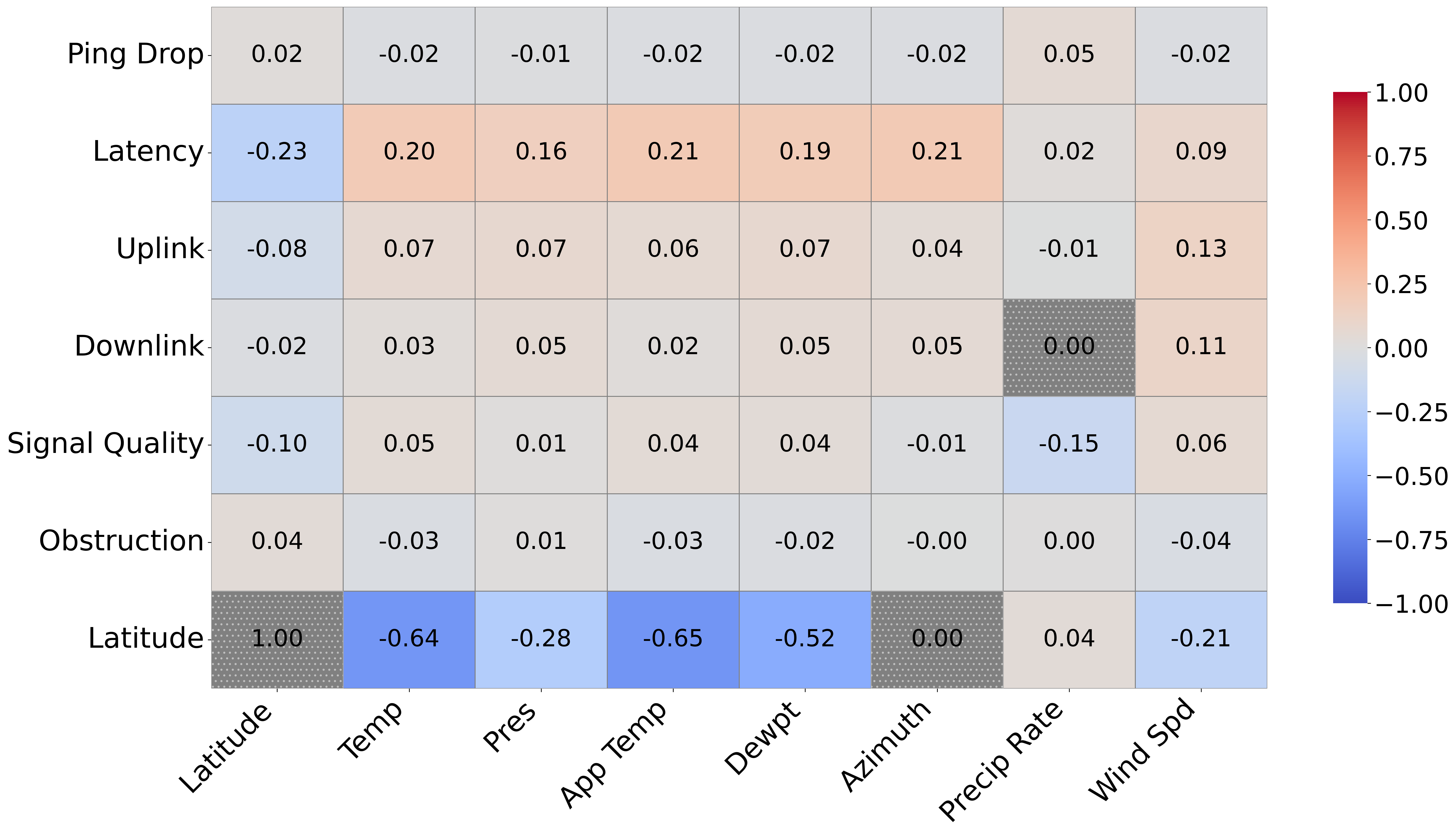}
 \caption{Pearson correlation matrix between key Starlink KPIs (rows) and selected environmental and geographic variables (columns). Latitude is also included as a final row to illustrate its significant correlations with several weather variables, highlighting potential confounding factors discussed in the text. Numeric values indicate the correlation coefficient ($r$), while hashed cells represent correlations that are not statistically significant ($p \ge 0.05$). Color intensity corresponds to the strength and direction of the significant correlations.}

 \label{fig:correlation_matrix}
\end{figure}

To understand the general relationships between environmental conditions and Starlink performance, Pearson correlation coefficients were calculated between key Starlink KPIs and selected continuous weather variables.  Figure \ref{fig:correlation_matrix} presents a heatmap of statistically significant Pearson correlations ($p < 0.05$) between key Starlink KPIs and selected weather variables that demonstrated a correlation of at least $|r| > 0.1$ with one or more Starlink KPIs.

The analysis reveals several statistically significant, albeit generally weak to moderate, linear correlations (Figure \ref{fig:correlation_matrix}). Notably, Ping Latency exhibits slight positive correlations with thermal metrics such as temperature ($r = 0.20$) and dew point ($r = 0.19$). A modest negative correlation exists between Ping Latency and geographic latitude ($r = -0.23$), suggesting a tendency for lower latencies at higher latitudes. In contrast, Signal Quality primarily shows a notable negative correlation with precipitation rate ($r = -0.15$). Ping Drop Rate does not demonstrate strong linear relationships with most environmental variables, with correlation coefficients below |r| = 0.1 . Throughput metrics (uplink and downlink) show some weak positive correlations with wind speed (e.g., $r = 0.13$ for uplink).

\begin{figure}[t]
  \centering
  \includegraphics[width=\linewidth]{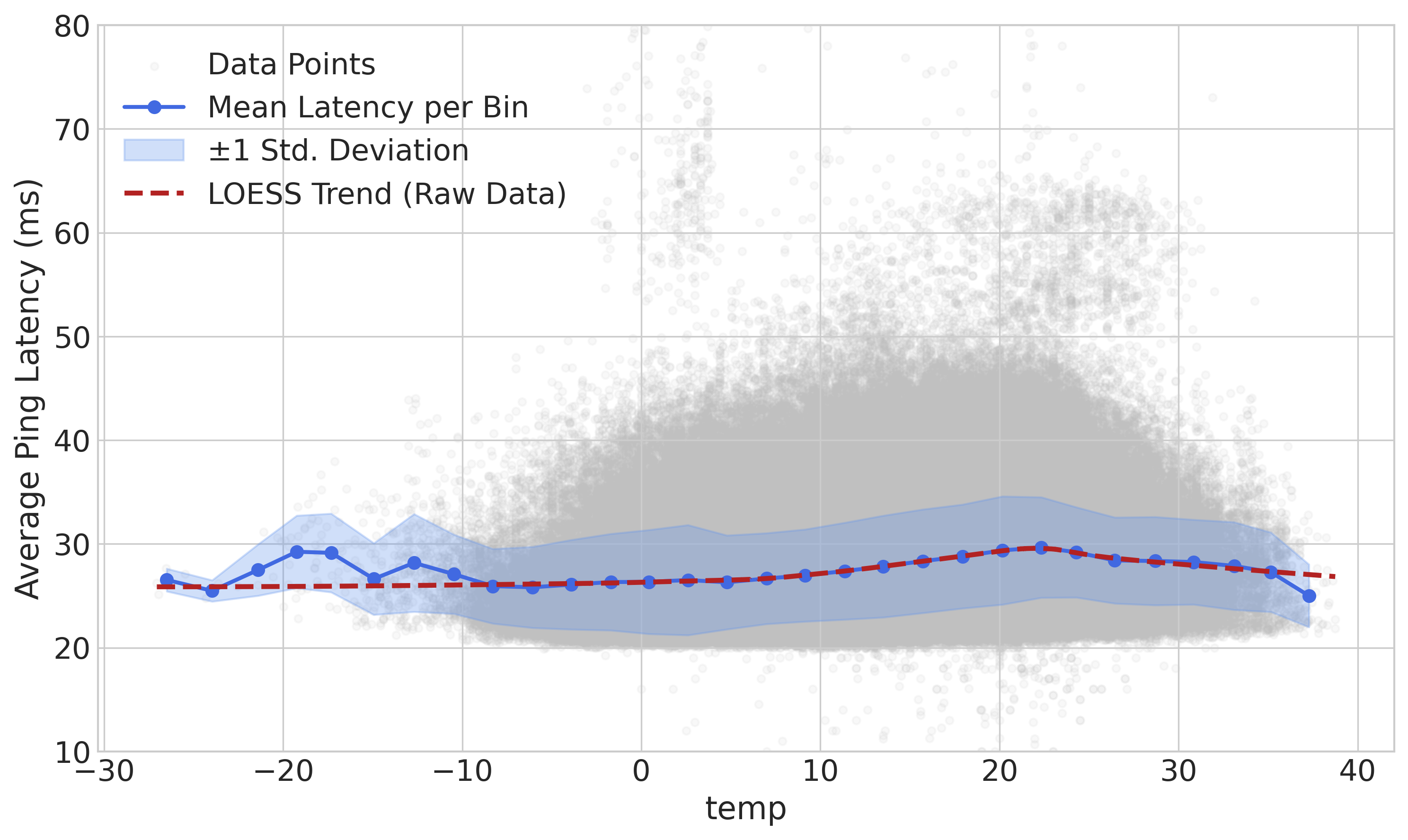} 
  \caption{Relationship between Temperature ($^{\circ}$C) and Average Hourly Ping Latency (ms). Each light point represents an hourly average for a terminal. The blue line indicates the mean latency calculated for temperature bins, with the shaded blue area representing $\pm$1 standard deviation around the binned means. The red line is a LOESS trendline, illustrating the general positive trend (overall Pearson $r = 0.20$). Despite the discernible upward trend with increasing temperature, average hourly latency predominantly remains well below 40 ms across the observed temperature range.}
  \label{fig:temp_latency_linear}
\end{figure}

Overall, while several environmental factors display statistically detectable linear relationships with Starlink KPIs, many of these, particularly those related to temperature and moisture, also correlate with latitude. This suggests that some observed weather-related correlations might be confounded by, or secondary to, factors related to geographic positioning and potentially Starlink's satellite constellation architecture relative to different latitudes. For instance, the positive linear correlation previously noted between ambient temperature and Ping Latency (Pearson $r = 0.20$) is explored in more detail in Figure \ref{fig:temp_latency_linear}. This figure, which includes LOESS trendline  (red) and binned averages (blue line with standard deviation shading), visually confirms a general upward trend in latency as temperature increases. However, the practical impact of this trend appears limited within the observed data range. Even at higher temperatures, the binned average latency, as well as the vast majority of individual hourly data points, predominantly remain well below 40 ms—a level generally associated with good user experience. This suggests that while ambient temperature has a statistically measurable relationship with latency, its magnitude within the typical operating conditions and temperature ranges encountered in this study does not appear to be a primary driver of severe, user-impactful latency degradation in isolation.

\subsection{Impact of Specific Weather Categories on Performance}
\label{ssec:results_weather_categories}
To assess the impact of different weather types more directly, particularly severe weather, key Starlink KPIs were aggregated and compared across various weather description categories obtained from the Weatherbit API. 

\begin{figure}[t] 
  \centering
  \includegraphics[width=\linewidth]{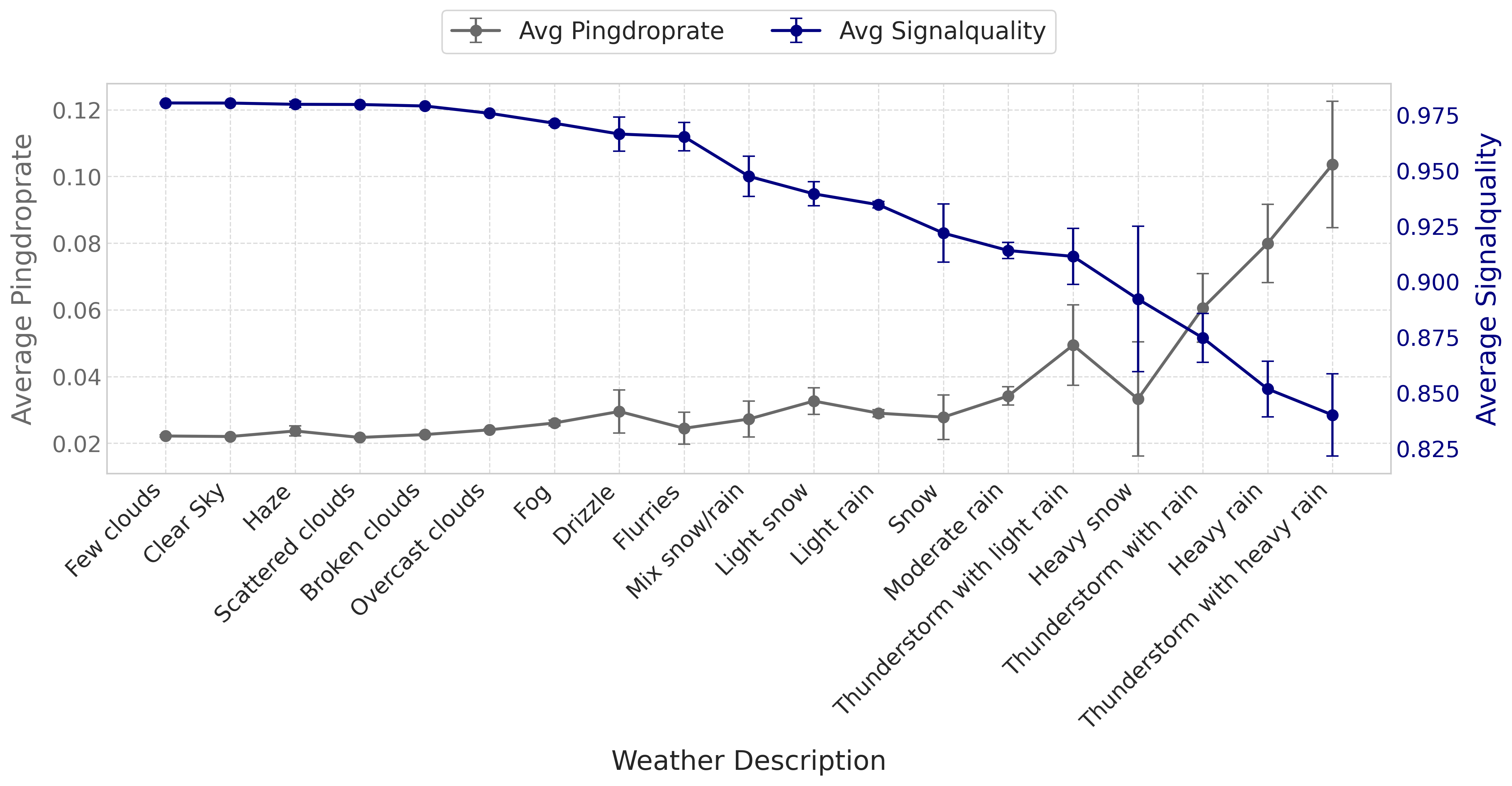} 
  \caption{Average Signal Quality (navy, left Y-axis, 0-1 scale, approx. range 0.82-0.98) and Average Ping Drop Rate (gray, right Y-axis, \%, approx. range 0.02-0.12) across different weather description categories, ordered by approximate severity. Points represent the mean for each category; error bars indicate 95\% confidence intervals for the mean. The diverging trends illustrate increasing performance degradation with weather severity.}
  \label{fig:kpi_trends_vs_weather_type}
\end{figure}

Figure \ref{fig:kpi_trends_vs_weather_type} simultaneously illustrates the trends in Average Signal Quality and Average Ping Drop Rate as weather conditions progress from clear skies to more severe precipitation-based events. A clear divergence is observable:
\begin{itemize}
    \item \textbf{Average Signal Quality (navy line):} This metric is highest during "Clear Sky" conditions (mean = 99) and systematically decreases with increasing weather severity. It reaches its lowest levels during events like "Thunderstorm with heavy rain" (mean = 84).
    \item \textbf{Average Ping Drop Rate (gray line):} Conversely, this metric is minimal during clear or mildly cloudy conditions (e.g., "Clear Sky" mean = 2\%) but increases substantially with weather severity, peaking during heavy precipitation and thunderstorms (e.g., "Thunderstorm with heavy rain" mean = 18\%).
\end{itemize}
These opposing trends, clearly depicted by the data points and their 95\% confidence intervals for the mean, strongly indicate that as weather conditions worsen, link quality degrades and data packet continuity is increasingly compromised.

Statistical analysis confirmed the observed trends. ANOVA indicated a highly significant effect of weather category on both \textit{Signal Quality} ($F(18, 799{,}575) = 1204.581$, $p < 0.001$) and \textit{Ping Drop Rate} ($F(18, 793{,}124) = 90.378$, $p < 0.001$). Post-hoc Tukey's HSD  tests ($p < 0.05$) further confirmed that performance during severe weather conditions, such as \textit{Thunderstorm with heavy rain}, was significantly worse---characterized by lower Signal Quality and higher Ping Drop Rate---when compared to mild conditions like \textit{Clear Sky} ($p_{\text{adj}} < 0.001$ for both KPIs). Several intermediate weather categories also exhibited statistically significant differences in performance, consistent with the degradation patterns illustrated in Figure~\ref{fig:kpi_trends_vs_weather_type}. These quantitative results support the statistical robustness of the observed weather-induced performance degradation.

\subsubsection{Likelihood of Significant Performance Issues}
\label{sssec:results_issue_likelihood}
Beyond average performance degradation, we analyzed the likelihood of terminals experiencing significant performance issues (defined as hourly average Signal Quality < 0.5 or Ping Drop Rate > 0.5, as detailed in Section \ref{ssec:definitions}) under different weather conditions. Overall, across all $\sim$870,000 terminal-hours in the dataset (encompassing both non-severe and severe weather), approximately 9.8\% of records exhibited at least one such performance issue, with the majority (90.2\%) operating without these defined significant degradations. 

However, this likelihood changes dramatically when isolating periods of severe weather. A Chi-squared test confirmed that the proportion of records with as issue is significantly higher during severe weather events compared to non-severe conditions ($\chi^2(1, N) = 3605.99$, $p < 0.001$). 

\begin{figure}[t] 
  \centering
  \includegraphics[width=\linewidth]{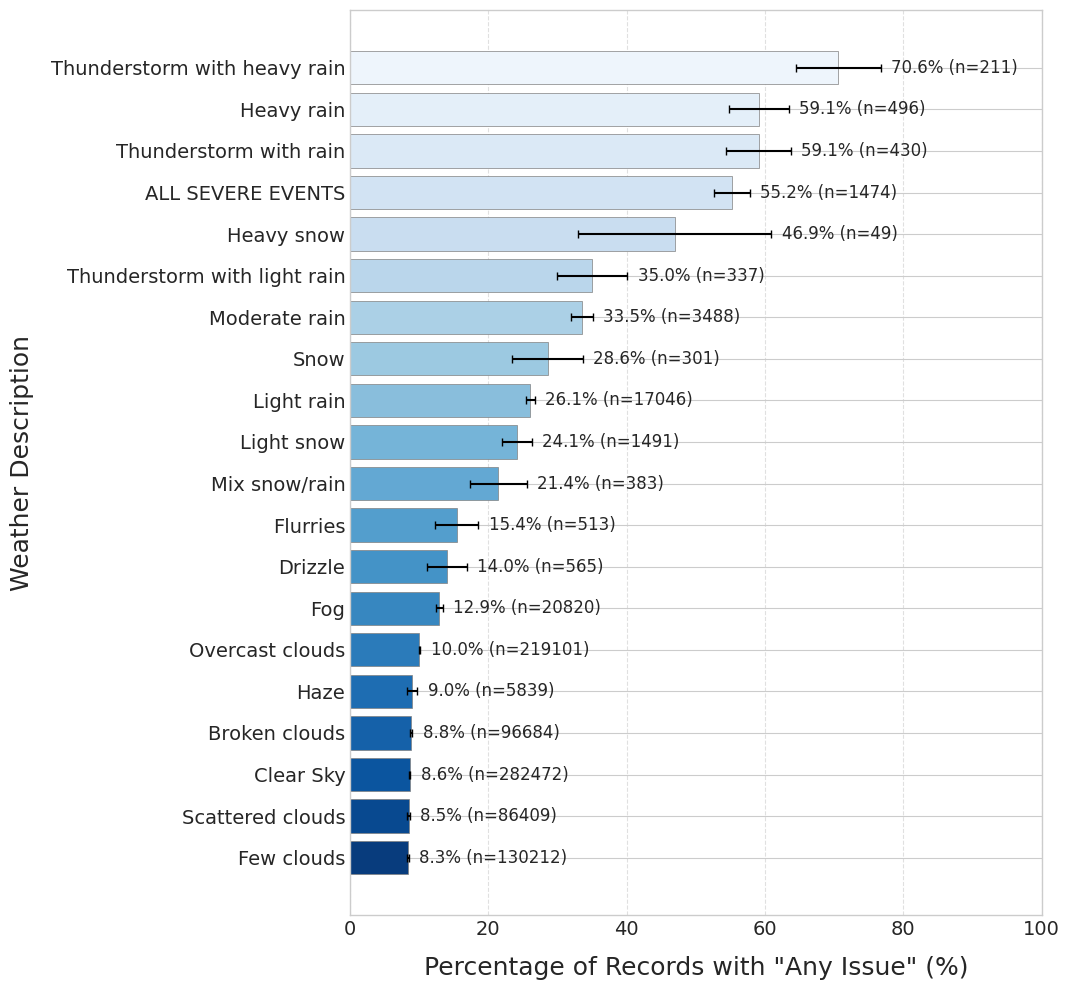} 
  \caption{Percentage of hourly records with significant performance issues (Signal Quality < 0.5 or Ping Drop Rate > 0.5) under different severe weather conditions. Percentages are labeled on top of each bar.}
  \label{fig:bar_severe_weather_issues}
\end{figure}

Table \ref{tab:severe_weather_impact_likelihood} details the percentage of hours within specific severe weather categories where performance issues were observed. Across all defined sever events (N=1,474 hours in this category), 55.2\% of records experienced at least one performance issue. This is substantially higher than the baseline issue rate observed across the entire dataset. Figure \ref{fig:bar_severe_weather_issues} visually reinforces this, illustrating the increased probability of issues across various severe weather types.

\begin{table*}[htbp]
\centering
\caption{Likelihood of Performance Issues during Severe Weather Events (Signal Quality < 0.5 or Ping Drop Rate > 0.5).}
\label{tab:severe_weather_impact_likelihood}
\begin{tabular}{@{}l r r r r r@{}}
\toprule
\textbf{Severe Weather Description} & \textbf{N hours} & \textbf{\% Any Issue} & \textbf{\% Low SQ} & \textbf{\% High PDR} & \textbf{\% Both} \\
\midrule
Thunderstorm w/ light rain & 337 & 35.0\% & 33.2\% & 13.6\% & 11.9\% \\
Sleet & 11 & 45.5\% & 36.4\% & 27.3\% & 18.2\% \\
Heavy Snow & 49 & 46.9\% & 42.9\% & 12.2\% & 8.2\% \\
Thunderstorm w/ rain & 430 & 59.1\% & 56.5\% & 24.4\% & 21.9\% \\
Heavy Rain & 496 & 59.1\% & 57.7\% & 30.8\% & 29.4\% \\
Thunderstorm w/ heavy rain & 211 & 70.6\% & 69.2\% & 40.3\% & 38.9\% \\
\midrule
\textbf{ALL SEVERE EVENTS} & \textbf{1,474} & \textbf{55.2\%} & \textbf{53.4\%} & \textbf{26.4\%} & \textbf{24.6\%} \\
\bottomrule
\end{tabular}
\end{table*}

As depicted in Figure \ref{fig:bar_severe_weather_issues} and detailed in Table \ref{tab:severe_weather_impact_likelihood}, "Thunderstorm with heavy rain" posed the greatest risk, with 70.6\% of such hours resulting in significant performance degradation. "Heavy rain" and "Thunderstorm with rain" also led to issues in approximately 59\% of recorded instances. Even "Thunderstorm with light rain" (35.0\%), "Sleet" (45.5\%), and "Heavy Snow" (46.9\%) demonstrated a substantially elevated likelihood of impacting service compared to benign conditions. These findings highlight that while severe weather events are less frequent, their capacity to disrupt Starlink service is considerably higher.

\subsection{Temporal Dynamics of Disruptions: Minute-Level Case Studies}
\label{ssec:results_deep_dives}
While aggregated hourly data reveals significant trends in performance degradation during severe weather, finer-grained temporal analysis is crucial to understand the precise duration and severity of user-experienced disruptions. Although our primary dataset consists of hourly aggregations, we also collect Starlink performance data at 15-second intervals, which we aggregate to minute-level resolution for detailed event analysis. The following case studies illustrate the value of this approach.

\subsubsection{Case Study 1: Service Degradation during Thunderstorm with Heavy Rain}
\begin{figure}[htbp]
  \centering
  \includegraphics[width=\linewidth]{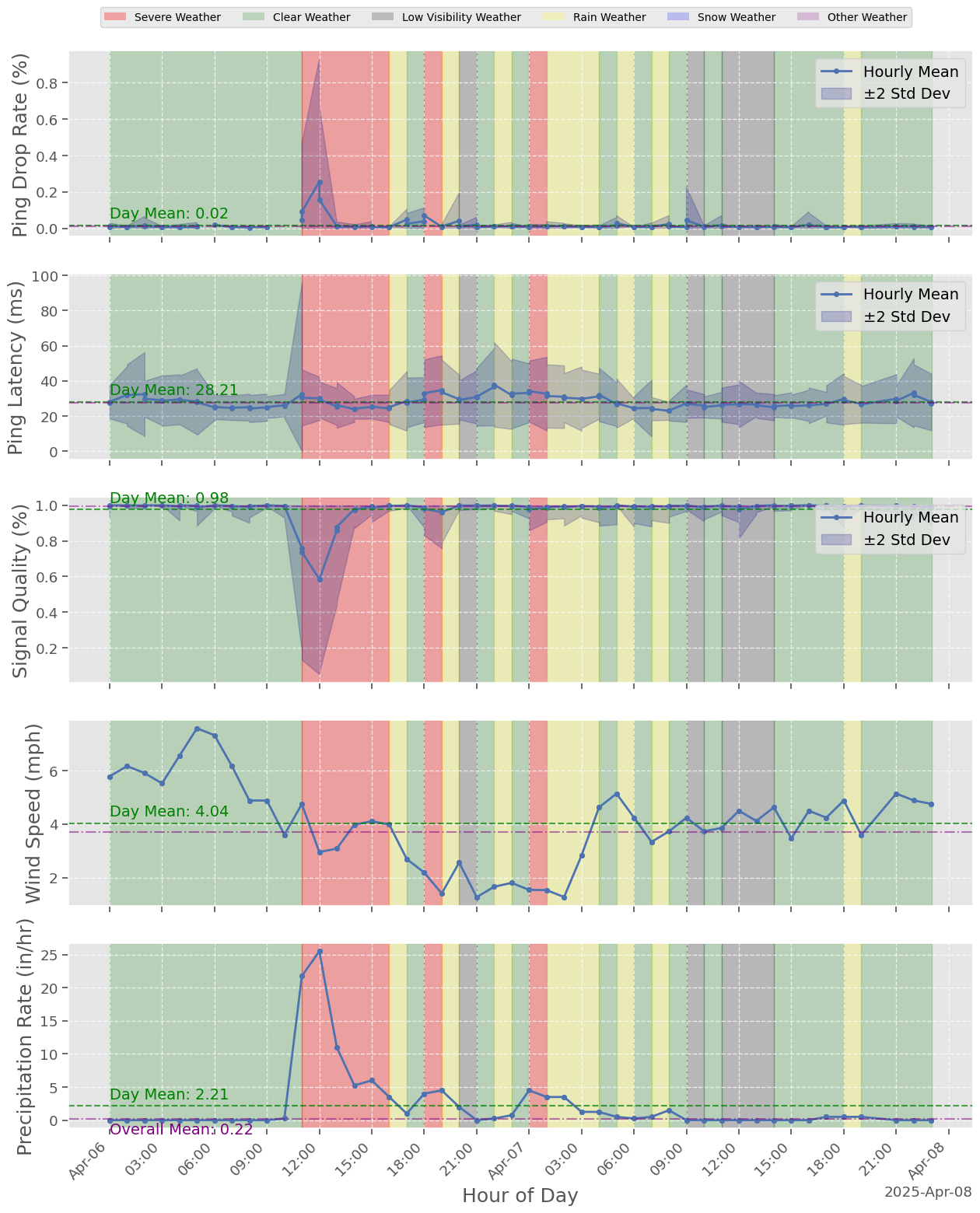} 
  \caption{Case Study 1 (Hourly Context): Hourly average Starlink performance metrics (e.g., Ping Drop Rate, Signal Quality) and relevant weather data (e.g., Precipitation Rate). The shaded in red region indicates the period identified as "Thunderstorm with heavy rain" at an hourly resolution.}
  \label{fig:case_study_1_hourly}
\end{figure}

Figure \ref{fig:case_study_1_hourly} presents an overview of hourly performance metrics for a representative Starlink terminal leading up to, during, and after a "Thunderstorm with heavy rain" event. A noticeable degradation, such as a peak in average hourly Ping Drop Rate to $0.22 \pm 0.45$ std and a dip in average hourly Signal Quality to $0.61 \pm 0.33$, is apparent during the hour(s) corresponding to the severe weather.

\begin{figure}[htbp]
  \centering
  \includegraphics[width=\linewidth]{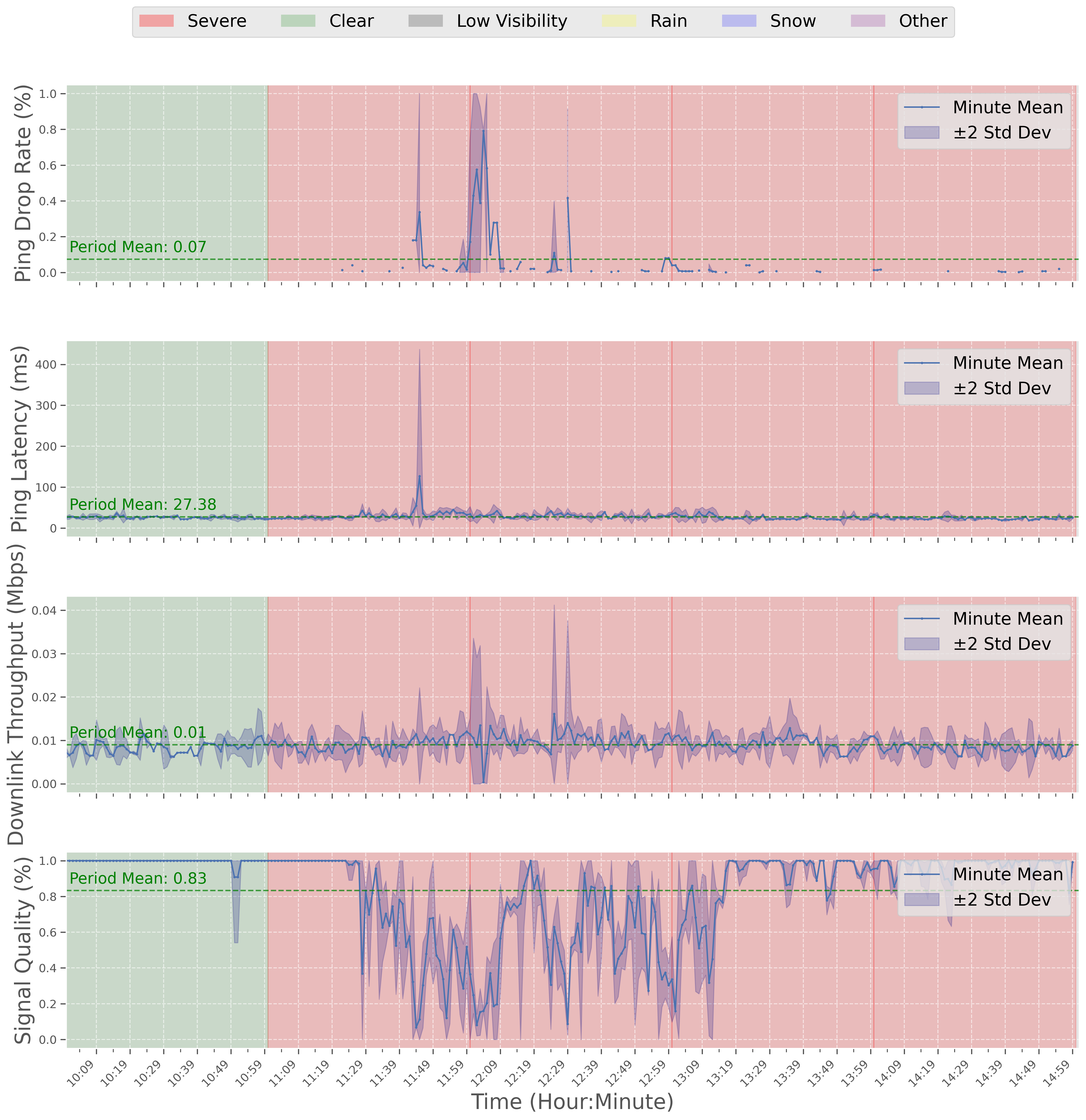} 
  \caption{Case Study 1 (Minute-Level Detail): Minute-level Signal Quality and Ping Drop Rate for the Starlink terminal shown in Fig. \ref{fig:case_study_1_hourly} during the "Thunderstorm with heavy rain" event. The highlighted in red corresponds to the period of most intense impact identified from hourly data, showing sustained degradation.}
\label{fig:case_study_1_minute}
\end{figure}

To investigate the precise nature of this degradation, Figure \ref{fig:case_study_1_minute} provides a minute-level view of Signal Quality, Ping Drop Rate, Ping Latency, and Downlink Throughput for the same event period. This granular analysis reveals that the hourly performance dip corresponded to a sustained period of approximately 1.5 hours where minute-level Signal Quality remained below 0.6 and Ping Drop Rate frequently spiked above 70-80\%. For instance, between 11:29 and 13:19, the link quality was severely compromised, illustrating how an hourly average can mask a period of near-unusable service.

\subsubsection{Case Study 2: Complete Service Outage during Thunderstorm with Heavy Rain}
\begin{figure}[htbp]
  \centering
  \includegraphics[width=\linewidth]{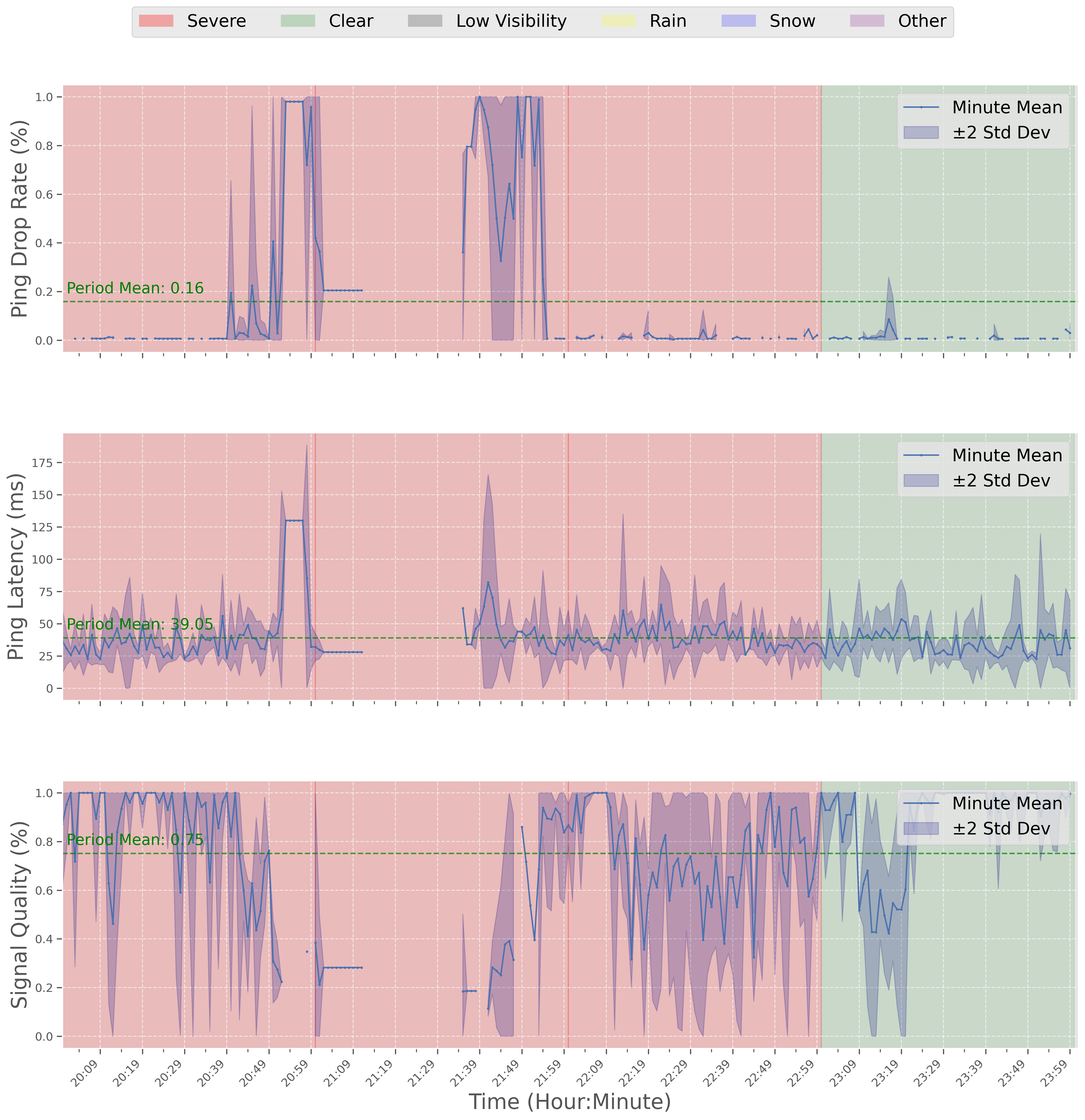} 
  \caption{Case Study 2: Minute-level performance metrics for a Starlink terminal during a "Thunderstorm with heavy rain" event, illustrating a period of complete service outage. Note the sustained period where the data gap accrues between 20:49-21:49.}
  \label{fig:case_study_2_minute}
\end{figure}

In more extreme instances, severe weather can lead to a complete, albeit temporary, loss of service. Figure \ref{fig:case_study_2_minute} presents minute-level performance data for a different Starlink terminal also experiencing a "Thunderstorm with heavy rain". During this event, a significant period of total service outage is evident. From approximately 20:49 to 21:49 data telemetry ceased, and Ping Drop Rate effectively reached 100\% prior to the outage, indicating a complete loss of connection for approximately 1 hour. This case highlights the capacity of severe weather events not only to degrade performance but also to cause temporary full service outages.

These case studies, representative of patterns observed during particularly intense weather conditions, underscore that while Starlink provides resilient service overall, acute severe weather episodes can lead to substantial, user-impactful disruptions whose true characteristics are best revealed through fine-grained temporal analysis.

\section{Discussion and Future Work}
\label{sec:discussion_future}

This paper presents a large-scale, spatio-temporal analysis of ~870,000 terminal-hours across 1,292 Starlink user terminals in the continental United States over a five-week period (Feb–Apr 2025). Our primary finding is that while the Starlink network exhibits high baseline stability---with over 90\% of terminal-hours showing no significant issues---this reliability is markedly challenged by severe weather. Events like thunderstorms with heavy rain caused performance degradation in over 70\% of affected hours, with minute-level case studies revealing that these disruptions can persist for minutes to over an hour, sometimes leading to full service outages.

These findings provide a quantitative foundation for developing the spatially-aware, weather-informed predictive models that are a key potential application of this work. For example, the quantified issue likelihoods presented in this study (Table \ref{tab:severe_weather_impact_likelihood}) can serve as empirical priors in machine learning models. By joining these historical patterns with gridded weather forecast products (e.g., HRRR), a spatio-temporal model could be trained to output a probabilistic forecast of service degradation for a given geographic area. Such a model could enable network operators to preemptively adjust routing, allow enterprise users to receive outage forecasts for mission-critical planning, and provide consumers with more realistic service expectations.

We acknowledge several limitations and confounding factors that represent important avenues for future research. Our analysis primarily correlates ground-level weather with performance and does not explicitly model key LEO network dynamics such as satellite elevation angle, where lower angles create a longer, more vulnerable atmospheric path, or inter-satellite handovers, both of which can independently affect performance. Furthermore, our study does not account for regional variations in satellite coverage density. While the physical principles of rain fade provide a strong causal basis for our primary findings, our observational approach establishes statistical correlation, and disentangling these operational factors from purely weather-induced effects is a complex challenge. Finally, while the proprietary nature of our dataset limits direct reproducibility, the methodology presented offers a generalizable framework for operators or researchers with access to similar telemetry.

Our future work aims to address these limitations directly. A key priority is to enrich our models by incorporating satellite ephemerides to calculate real-time elevation angles and better approximate atmospheric path conditions. We also plan to systematically quantify the duration, frequency, and recovery patterns of performance disruptions across the full dataset, moving beyond the illustrative case studies presented here. Finally, expanding the analysis geographically and temporally is critical for building generalizable models that account for diverse global regions, seasons, and the impact of different Starlink hardware and firmware revisions.

We view this work as a foundational step toward weather-resilient LEO networking, where real-time environmental intelligence is integrated into the performance management and orchestration layers of satellite internet infrastructure. Such integration is essential for ensuring reliable connectivity in dynamic atmospheric conditions—particularly in the most remote and underserved regions where LEO systems offer the greatest transformative potential.

\begin{acks}
The authors would like to thank the entire engineering and data platform teams at Armada for their support in developing and maintaining the infrastructure that made this research possible. We extend special gratitude to Vladimir Yakovlev for his invaluable assistance with the data platform and in facilitating access to the high-resolution performance telemetry data that forms the foundation of this study. We are also grateful to the anonymous reviewers for their insightful feedback, which significantly improved the quality of this paper.
\end{acks}

\bibliographystyle{ACM-Reference-Format}
\bibliography{ACMSigspatial_LEO}


\end{document}